\documentclass[letterpaper]{article} % DO NOT CHANGE THIS
\usepackage{aaai2026}  % DO NOT CHANGE THIS
\usepackage{times}  % DO NOT CHANGE THIS
\usepackage{helvet}  % DO NOT CHANGE THIS
\usepackage{courier}  % DO NOT CHANGE THIS
\usepackage[hyphens]{url}  % DO NOT CHANGE THIS
\usepackage{graphicx} % DO NOT CHANGE THIS
\usepackage{natbib}  % DO NOT CHANGE THIS AND DO NOT ADD ANY OPTIONS TO IT
\usepackage{caption} % DO NOT CHANGE THIS AND DO NOT ADD ANY OPTIONS TO IT
\usepackage{algorithm}
\usepackage{algorithmic}

\usepackage{newfloat}
\usepackage{listings}
\DeclareCaptionStyle{ruled}{labelfont=normalfont,labelsep=colon,strut=off} % DO NOT CHANGE THIS
\floatstyle{ruled}
\newfloat{listing}{tb}{lst}{}
\floatname{listing}{Listing}
\usepackage{tikz}
\usetikzlibrary{shapes, backgrounds}
\usepackage{enumitem}
\usepackage{amsmath}
\usepackage{booktabs}
\usepackage{arydshln}
\usepackage{multirow}
\usepackage{comment}
\usepackage{amsfonts}

\newcommand{\revision}[1]{\textcolor{black}{#1}}
\newcommand{\rev}[1]{\textcolor{black}{#1}}

\newcommand{\revAIES}[1]{\textcolor{black}{#1}}

\copyrightyear{2026}
\nocopyright 

\title{\revision{Small-world Networks of Agents Brainstorm AI Risks}}

\author{
    Ke Zhou\textsuperscript{\rm 1,2}
    Edyta Bogucka\textsuperscript{\rm 1},
    Daniele Quercia\textsuperscript{\rm 1,3}\\
}

\affiliations{
    \textsuperscript{\rm 1}Nokia Bell Labs, UK\\
    \textsuperscript{\rm 2}University of Nottingham, UK\\
    \textsuperscript{\rm 3}Politecnico di Torino, Italy\\
    ke.zhou@nokia-bell-labs.com, edyta.bogucka@nokia-bell-labs.com, quercia@cantab.net
}

\begin{document}

\maketitle

\begin{abstract}

\revAIES{The ideation phase of participatory AI risk assessment often starts with a blank slate or a limited list of predefined risks, making it difficult to surface indirect or systemic harms.}
\revAIES{To address this limitation, we propose a three-stage tool to support the ideation phase of AI risk assessment. The tool complements participatory AI, rather than replacing it, and helps focus later engagement with affected communities.}
First, it dynamically discovers stakeholders depending on the given AI use and recursively expanding outward, allowing overlooked or indirect stakeholders to emerge. Second, it simulates these stakeholders with LLMs (creating in-silico stakeholders), connecting them into a network of a given topology, and having them ideate about risks. Third, it prioritizes risks using network centrality measures (the centrality of in-silico stakeholders in the network). In an initial evaluation, we found that betweenness centrality run through agents connected in a small-world network works best as it elevates risks raised by stakeholders who bridge disconnected groups, surfacing novel, systemic harms that traditional methods often miss. On an AI chatbot companion use case, this approach increased the novelty of the identified risks by approximately 1.1 points over single LLM brainstorming, and by 0.5 points over agentic LLM brainstorming, measured on a normalized five-point Likert scale, without reducing the plausibility or severity of the identified risks. To test whether our framework helps a human-led ideation session using the Futures-Wheel approach, we divided 11 teams of non-western young chatbot users into two types: control (team) and treatment (team) in a participatory AI risk assessment. The control teams started from a list of risks generated by the 45 AI practitioners in the initial evaluation; the treatment teams started from a list generated by our framework. The treatment teams identified more risks overall, and more systemic, human-computer interaction, and environmental risks in particular, categories that practitioners typically find harder to spot. 
%Supplementary materials and  results are available at \url{https://social-dynamics.net/ai-risks/small-world}.
\end{abstract}

\begin{links}
    \link{Project}{https://social-dynamics.net/ai-risks/small-world}
\end{links}

\section{Introduction}

\revision{As AI systems are increasingly integrated into critical societal infrastructure, from healthcare diagnostics to companion chatbots for vulnerable populations, the imperative for comprehensive impact and risk assessment has transitioned from a theoretical ideal to a governance necessity. A critical but often under-examined stage in this process is \emph{risk ideation}: the initial step in which participants attempt to enumerate as many relevant risks as possible before prioritization and mitigation. Because all downstream analysis depends on this initial pool, the quality and breadth of ideation largely determine the effectiveness of the entire assessment.}
\revision{In practice, however, ideation frequently begins from a blank slate or a limited, organizer-defined list of risks. This creates a structural bottleneck. While frameworks like the EU AI Act \cite{EUACT2024} emphasize the need to catch ``systemic risks'', participants tend to anchor on familiar, high-frequency risks (e.g., privacy or security), while overlooking indirect, systemic, or stakeholder-specific harms \cite{lancaster2024s, xia2023towards, frohling2026agent}.} \revision{This limitation is not merely a matter of scale, but of epistemic diversity: risks that emerge from the interaction between stakeholders, or that disproportionately affect marginalized groups, are systematically underrepresented \cite{birhane2022power, lancaster2024s}.} 
%, vecchione2021algorithmic

Traditional human workshops offer depth but are unscalable and constrained by the ``bounded rationality'' of small groups of participants \cite{diehl_stroebe_productivity_loss_brainstorming87}. Conversely, automated tools that use Large Language Models (LLMs) to generate risk lists, such as Farsight \cite{wang2024farsight}, ExploreGen \cite{herdel_et_al_exploregen25}, and AHA! \cite{buçinca_aha_generating_ai_risks23}, act as single expert oracles. They tend to converge on ``consensus'' risks: those that are plausible and severe, but often generic. Consequently, they struggle to explore the long tail of socio-technical failures that only emerge through the friction of conflicting perspectives, i.e., risks that %while statistically rare, 
often carry systemic consequences \cite{frohling2026agent} (e.g., developing an unhealthy over-reliance on the chatbot that hinders real-life social relationship development). In short, current methods are good at confirming what we already know (high likelihood), but less effective at discovering what we do not (high novelty). \revision{As a result, they fail to adequately support the human ideation process.}

\begin{figure*}[t]
    \centering
    \includegraphics[width=\linewidth]{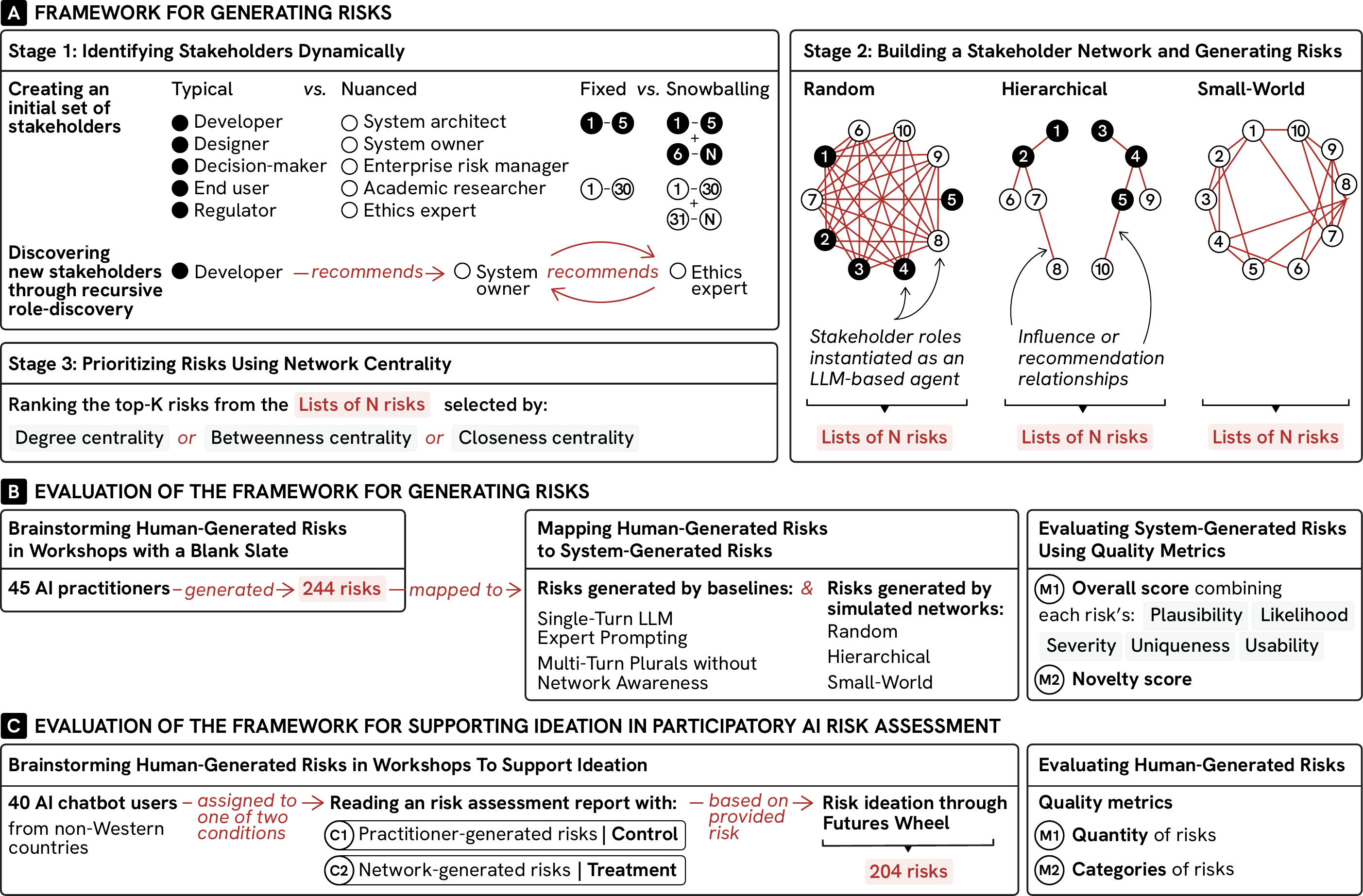}
    \caption{\revision{\textbf{Overview of the proposed framework and its two evaluations.} (A) Our three-stage framework for generating risks. It involves identifying diverse stakeholders dynamically (Stage 1); building a stakeholder of in-silico agents network and generating risks by orchestrating the communication through a network with a given topology (Stage 2); and prioritizing risks using three centrality metrics: degree, betweenness, and closeness (Stage 3). (B) Evaluation of the framework's risk generation quality by comparing system-generated risks with practitioners-generated risks from workshops. (C) Evaluation of the framework as a support tool for participatory AI risk assessment. Participants are assigned to either a control condition (report with practitioners-generated risks) or a treatment condition (report with network-generated risks), and use the provided reports as starting points for Futures Wheel ideation.}}
    \label{fig:methodology}
\end{figure*}

To address this, \revAIES{we propose an ideation tool designed specifically to overcome the ``blank slate'' bottleneck in AI risk assessment. The tool seeds brainstorming workshops with a network of LLM agents, surfacing more diverse and systemic risks than human sessions seeded with practitioner-generated risks.}
We make three main contributions:
\bigskip
\bigskip

\textbf{(1)} \emph{A Three-Stage Framework for Networked Stakeholder Simulation 
    (Section \ref{sec:networkagent}):} We propose a framework that dynamically discovers stakeholders via recursive snowballing, instantiates them as in-silico LLM agents, and connects them into a network topology through which they communicate %and deliberate 
    to brainstorm risks. 
    Risks are then prioritized using network centrality measures applied to the resulting stakeholder graph. We show small-world topologies combined with betweenness centrality are uniquely effective at surfacing novel, systemic harms that flat ideation methods miss.

\textbf{(2)}     \revision{\emph{Evaluation of Risk Generation Quality (Section \ref{sec:eval1}):} We evaluate our framework against single-LLM and multi-agent baselines using a dataset of 244 risks generated by 45 AI practitioners in blank-slate brainstorming workshops. Our approach improves novelty by approximately 1.1 points over single-LLM brainstorming, and 0.5 points over agentic LLM baselines, without reducing plausibility or severity.}

\textbf{(3)}     \revision{\emph{Evaluation of Support for Human-Led Participatory AI Risk Assessment (Section \ref{sec:futureswheel}):} We conduct a controlled Futures Wheel study \cite{glenn_futures_wheel} with 40 non-Western young chatbot users. Treatment teams seeded with our framework's risks identified substantially more risks overall, and significantly more systemic, human-computer interaction, and socioeconomic harms than control teams seeded with practitioner-generated risks.}

\section{Related Work}
\begin{comment}
Our work builds on and connects several research strands: AI risk identification and assessment (Section \ref{sec:risk}), AI-augmented brainstorming (Section \ref{sec:brainstorm}), 
and the application of network science to stakeholder analysis (Section \ref{sec:networkscience}). 
\end{comment}

\subsection{AI Risk Identification and Assessment}
\label{sec:risk}

AI risk identification and AI risk assessment are distinct but related processes: AI risk identification uncovers potential harms, while risk assessment quantifies them for prioritization and decision-making. AI risk identification has evolved from narrowly defined technical safety evaluations toward broader socio-technical analyses. Early approaches surfaced technical risks at the model level, such as robustness failures and security vulnerabilities, and quantified them using metrics including error rates, perturbation robustness, and adversarial vulnerability \cite{schmitz2025global, steimers2022sources}. As AI systems became embedded in social and institutional contexts, research attention shifted toward identifying and quantifying downstream harms.

AI risk taxonomies have played a central role in this shift. Influential work by~\citeauthor{weidinger2022taxonomy} (2022), alongside subsequent syntheses \cite{uuk2024taxonomy, zhang2025dark}, provides comprehensive catalogues of risks spanning technical failures, human–AI interaction risks, and long-term systemic societal risks. These taxonomies offer shared vocabulary and conceptual clarity, but they are largely retrospective: they systematize known risks rather than supporting the anticipation of novel, context-specific risks in emerging applications. Moreover, they typically abstract away from the perspectives of distinct stakeholders.

Participatory approaches attempt to address these limitations by incorporating diverse human perspectives into risk identification \cite{delgado2023participatory, lee2019webuildai}. Methods such as the \emph{Judgment Call} card game \cite{ballard_et_al_judgment_call_game19}, participatory scenario planning \cite{kieslich2025scenario}, and end users auditing \cite{WeAudit2025} leverage structured interaction to surface values, concerns, and experiential knowledge that may otherwise remain implicit. \revision{While effective in eliciting rich insights, these approaches are resource-intensive, difficult to scale, and constrained by practical barriers to participation. In particular, assembling representative groups of end users, regulators, and marginalized stakeholders is often infeasible for organizations conducting AI risk assessments. In addition, current participatory AI literature \cite{kallina2025stakeholder, birhane2022power, corbett2023power, delgado2023participatory} overlooks the ``blank slate'' bottleneck in risk ideation, where forcing participants to brainstorm from scratch exhausts their cognitive bandwidth on obvious issues \cite{hohma2023investigating}.}

Current AI risk assessment practices are strongly influenced by regulatory frameworks \cite{EUACT2024}, and predominantly focus on evaluating identified risks in terms of likelihood and severity, an approach inherited from risk assessment traditions in domains such as safety engineering and finance. 
%These dimensions also feature prominently in the research literature as target evaluation metrics \cite{wang2024farsight}. 
These dimensions are also prominent evaluation criteria in prior research \cite{wang2024farsight}.
By contrast, novelty and diversity of perspectives have rarely been treated as explicit assessment objectives \cite{herdel_et_al_exploregen25}, and there remains a lack of robust measures for how they should be defined and operationalized. %For example, ISO/IEC 42005 AI Impact Assessment standard articulates five Responsible AI pillars (security and privacy, accountability, reliability, transparency, and fairness) as potential dimensions of diversity \cite{iso2025aiSystemImpactAssessment}.

\subsection{AI-Augmented Brainstorming for Risk Assessment}
\label{sec:brainstorm}

\revision{The use of AI systems to support brainstorming and ideation has gained increasing attention \cite{wang2025aideation, ashkinaze2025ai, choi2024creativeconnect}. Empirical studies show that LLMs can substantially increase the quantity of generated ideas \cite{bouschery_et_al_ai_augmented_brainstorming24, yu_han_et_al_llm_chatbot_human_number_creativity23}.} \revision{However, these gains are frequently accompanied by reductions in diversity and originality, with models exhibiting fixation on stereotypical or high-frequency patterns \cite{wadinambiarachchi_ai_design_fixation24, meincke2025chatgpt, AlDahoul2025, bai2025explicitly}.
This tendency is particularly salient in risk assessment contexts. For example, when prompted to identify ``AI risks'', LLMs often default to generic concerns such as privacy or bias, overlooking subtle, domain-specific, or interactional harms \cite{lancaster2024s, xia2023towards, frohling2026agent}.} 

\revision{To scale risk discovery, the industry has increasingly turned to AI red teaming and AI-augmented brainstorming. Red teaming leverages adversarial testing to probe models for vulnerabilities and policy violations \cite{ge2024mart, ganguli2022red}. While automated red teaming is highly effective at identifying technical exploits (e.g., jailbreaks), it fundamentally relies on predefined threat models and often struggles to anticipate nuanced, socio-technical harms that only emerge through complex stakeholder interactions.}
\revision{To broaden the scope of risk identification beyond red teaming,} several tools leverage LLMs to assist AI risk assessment directly. Systems such as Farsight \cite{wang2024farsight}, ExploreGen \cite{herdel_et_al_exploregen25}, and AHA! \cite{buçinca_aha_generating_ai_risks23} generate or visualize candidate risks based on system descriptions. While these tools improve efficiency, they typically %adopt a ``human-consults-AI'' paradigm and 
treat the LLM as a single expert oracle. As a result, they do not explicitly model interaction, disagreement, or perspective-taking among stakeholders, mechanisms that are known to underpin diversity in human group brainstorming \cite{meincke2025chatgpt}.

\subsection{Stakeholder Analysis and Network Science}
\label{sec:networkscience}

Recent work has explored the use of LLMs to simulate aspects of human behavior and social interaction, often described as \emph{in-silico} social science. The Plurals framework \cite{ashkinaze2025plurals} demonstrates that LLM agents initialized with distinct roles or personas can participate in structured deliberative processes, producing outputs that differ meaningfully from single-agent prompting. Similarly, Park et al.'s \emph{Generative Agents} \cite{park2023generative} show that agents endowed with memory and social context can exhibit emergent social dynamics.
Within AI risk identification, however, prior applications of multi-agent LLM systems have largely relied on unstructured debate or flat ideation structures. \citet{llmmultiagent} explored agent-supported brainstorming, and found that benefits were primarily limited to general or non-specialized topics. These findings suggest that role differentiation alone may be insufficient; the structure of interaction itself plays a critical role in shaping outcomes.

Stakeholder theory posits that organizational outcomes depend on the relationships among stakeholders, not solely on individual attributes \cite{parmar2010stakeholder}. Network science provides formal tools for analyzing these relationships. Rowley \cite{rowley1997moving} argues that stakeholder influence is a function of network position such as centrality or brokerage rather than intrinsic importance alone.
In adjacent domains, network measures \cite{borgatti1998network, watts1998collective} have been used to identify influential actors, information bottlenecks, and structural vulnerabilities. However, these methods have rarely been applied to in-silico brainstorming or assessment processes. Existing approaches typically aggregate risks by frequency or expert judgment, implicitly assuming all contributors are equally informative. %\todo{network in information diffusion and citations}

\smallskip
\noindent \textbf{Research Gap.} The initial ideation phase often starts from a ``blank slate'' or a narrow set of predefined risks. This forces participants to spend time listing well-known concerns before they can address broader, systemic risks. 
To address this gap, we introduce a tool that helps generate early ideas about AI risks by simulating discussions among in-silico stakeholders. The tool structures these interactions as networks, allowing us to explore how communication patterns shape the risks they identify. This approach surfaces risks from a wider range of perspectives, especially from stakeholders who ``bridge'' otherwise separate groups.

\section{Our Framework}
\label{sec:networkagent}

\revAIES{
Our framework operationalizes AI risk brainstorming through network-mediated ideation, a process where in-silico agents representing diverse stakeholders generate risks within a network topology.
Rather than treating risk identification as an isolated ideation task, we conceptualize it as a collaborative activity shaped by who participates, how they are connected, and whose perspectives bridge across roles.}

Figure~\ref{fig:methodology} provides an overview of the methodology, consisting of three stages: 
(1) Discovering stakeholders dynamically (Section \ref{sec:stakeholders});
(2) Building a stakeholder network (Section \ref{sec:network});
and (3) Prioritizing risks using network centrality (Section \ref{sec:centrality}).
Formally, given an AI use case, our goal is to produce a set of top-$K$ risks
$\mathcal{R} = \{r_1, r_2, \dots, r_K\}$ on a ranked list,
where each risk $r_k$ is associated with one or more stakeholders.

\subsection{Stage 1: Identifying Stakeholders Dynamically}
\label{sec:stakeholders}

\noindent \textbf{Creating An Initial Set of Stakeholders.}
We begin by identifying an initial set of stakeholders relevant to the AI use case $U$:
%\[
$\mathcal{S}_0 = \{s_1, s_2, \dots, s_m\}$,
%\]
where each $s_i$ corresponds to an abstract role rather than a specific individual. 

We define a \emph{typical stakeholder set} grounded in responsible AI practice and governance literature \cite{san2025ensuring}. This set includes five roles: \emph{Developers}, responsible for building and maintaining the AI system; \emph{Designers}, shaping user interaction and experience; \emph{Decision-makers}, responsible for deployment and organizational strategy; \emph{End users}, who directly interact with or are affected by the system; and \emph{Regulators}, responsible for oversight and compliance.
These roles align with prior work on AI auditing and accountability \cite{san2025ensuring}, which emphasizes the inclusion of decision subjects, domain experts, and regulatory actors in assessment processes. For AI companion chatbots in particular, these categories capture both technical and socio-emotional dimensions of risk.

We also define a \emph{nuanced stakeholder set} based on NIST standard framework for AI risk assessments \cite{ai2023artificial}, including AI developers, machine learning engineers, data scientists, data collectors and curators, system architects, model trainers, and technical auditors; product and project managers, system owners, organizational leadership, enterprise risk managers, legal counsel, compliance officers, procurement teams, and IT deployment and operations staff; clinicians, judges, HR professionals, and analysts; external regulators, supervisory authorities, policymakers, and legislators; and academic researchers, ethics experts, journalists and environmental stakeholders.

\smallskip
\noindent \textbf{Making The Set of Stakeholders Comprehensive Via Snowballing.}
To capture stakeholders omitted by the initial abstraction, we perform iterative stakeholder expansion by employing an LLM-driven snowballing technique for recommending more relevant stakeholders recursively. 
At iteration $t$, each stakeholder $s_i \in \mathcal{S}_{t}$ proposes a set of additional stakeholders
%\[
$\Delta \mathcal{S}_i^{(t)} = \{s_{i1}, s_{i2}, \dots\}$,
%\]
where we query the LLM : \textit{``Given the AI use case [Description] and your stakeholder role as [Stakeholder], list what other stakeholders (no personal names) are significantly impacted by or have influence over this system?''}
The candidate set for iteration $t+1$ is:
%\[
$\mathcal{S}_{t+1} = \mathcal{S}_t 
\bigcup_i 
\Delta \mathcal{S}_i^{(t)}$.
%\]

To prevent unbounded growth, we deduplicate roles using semantic similarity. Let $\phi(s)$ denote an embedding of a stakeholder description. A newly proposed stakeholder $s'$ is merged with an existing stakeholder $s$ if:
%\[
$\cos(\phi(s'), \phi(s)) \ge \tau$,
%\]
%
Expansion terminates when either there are no novel stakeholders are added ($\mathcal{S}_{t+1} = \mathcal{S}_t$), or a maximum number of iterations $T_{\max}$ is reached.

\subsection{Stage 2: Building a Stakeholder Network and Generating Risks}
\label{sec:network}

\noindent \textbf{Connecting Stakeholders In A Network Within A Given Topology.}
We define a stakeholder interaction network as a directed weighted graph:
%\[
$G = (V, E, W)$,
%\]
where: $V = \mathcal{S}_T$ is the final stakeholder set, $E \subseteq V \times V$ represents recommendation of stakeholders indicting influences, $W: E \rightarrow \mathbb{R}^+$ assigns edge weights.
An edge $(s_i, s_j)$ exists, if stakeholder $s_j$ is recommended by stakeholder $s_i$, and therefore deemed to maintain influence over stakeholder $s_j$ during risk ideation.
Using this stakeholder set, we generate multiple instantiations of $G$ with different network topologies:

%\paragraph{
\emph{Hierarchical Network:}
It is constructed by assigning stakeholders to levels $\ell(s) \in \mathbb{N}$ based on organizational authority. Edges are constrained such that:
%\[
$\ell(s_i) < \ell(s_j) \Rightarrow (s_i \rightarrow s_j)$.
%\]
This reflects top-down flow typical of formal governance structures.

%\paragraph{
\emph{Small-World Network:} 
It is designed to balance dense local interaction among closely related stakeholder roles with a limited number of long-range connections across the network. We construct a small-world--inspired directed acyclic graph by preserving strong high-weight edges within stakeholder clusters identified through the recommendation, while enforcing acyclicity. Formally, an edge is retained between stakeholders $s_i, s_j \in V$, if one recommends the other, and the edge does not form circles. This structure supports locally coherent ideation, while enabling risks to propagate across stakeholder communities through bridging roles.

%\paragraph{
\emph{Random Network:}
In the random baseline, edges are sampled uniformly at random from $V \times V$, subject to acyclicity and fixed edge count $|E|$.

\noindent \textbf{Pruning Network Graph and Removing Cycles.}
Given the directed interaction structures, we transform $G$ into a directed weighted graph $G'$ by performing backbone extraction, and graph cycle removal procedures.

\noindent \emph{Backbone Extraction.}
The initial stakeholder interaction graph $G = (V, E, W)$ may contain a dense set of edges reflecting weak, redundant, or noisy relationships. We apply a disparity filter \cite{serrano2009extracting} to remove statistically insignificant edges, retaining only edges whose normalized weight exceeds a significance threshold $\delta$.
We set $\delta = 0.05$ following prior work on backbone extraction in weighted networks \cite{serrano2009extracting}.
Applying the disparity filter yields a pruned graph:
%\[
$G_b = (V, E_b, W_b)$,
%\]
where $E_b \subseteq E$ contains only statistically significant edges. This backbone graph retains the majority of total edge weight, while substantially improving interpretability and computational efficiency.

\noindent \emph{Graph Cycle Removal.}
The network-mediated ideation structures are required to be directed acyclic graphs (DAGs) in order to support multi-round, feed-forward information flow. However, the backbone graph $G_b$ may still contain cycles, reflecting reciprocal or mutual influence among stakeholders. To enforce acyclicity, we therefore transform $G_b$ into a DAG $G'$ using a greedy maximum-weight acyclic subgraph construction.
Given a directed weighted graph $G_b = (V, E_b, W_b)$, we seek a directed acyclic subgraph $G' = (V, E', W')$ such that: $E' \subseteq E_b$,
and the total retained weight $\sum_{(i,j) \in E'} w_{ij}$ is maximized.
This problem is NP-hard in general. We therefore adopt an efficient greedy approximation: sorting remaining edges by descending weight,  and greedily adding them to $G'$ unless doing so would create a cycle.
This algorithm prioritizes preserving high-weight 
influence relationships while eliminating weaker reciprocal ties that would violate acyclicity. As a result, $G'$ approximates the most influential pathways within the stakeholder system. To assess robustness, we verified that alternative cycle-breaking strategies (e.g., random edge removal) produce qualitatively similar results in downstream risk diversity and ranking, though with slightly higher variance. %We therefore adopt the greedy maximum-weight approach as a principled balance between fidelity and stability.

%\subsubsection{Generating Risks By Orchestrating The Communication Among Stakeholders.}
\smallskip
\noindent \textbf{Generating Risks By Orchestrating The Communication Among Stakeholders.}
Each stakeholder $s_i \in V$ is instantiated as an in-silico agent $a_i$, parameterized by:
%\[
$a_i = \langle s_i, U, \mathcal{I}_i^{(t)} \rangle$,
%\]
where $\mathcal{I}_i^{(t)}$ denotes information available at round $t$ for the use case $U$.
Stakeholder agents generate risks via conditional language modeling:
%\[
$r_{ik}^{(t)} \sim p_\theta(r \mid s_i, U, \mathcal{I}_i^{(t)})$,
%\]
where $\theta$ denotes the underlying LLM parameters.
At each round $t$, stakeholder agent $a_i$ receives inputs from its immediate predecessors:
%\[
$\mathcal{I}_i^{(t)} = \bigcup_{(s_j \rightarrow s_i) \in E'} \mathcal{R}_j^{(t-1)}$,
%\]
where $\mathcal{R}_j^{(t-1)}$ is the set of risks generated by stakeholder agent $a_j$ in the previous round.
This induces a structured dependency between stakeholder agents’ outputs, operationalizing network-mediated ideation.

\subsection{Stage 3: Prioritizing Risks Using Network Centrality}
\label{sec:centrality}

Let $G' = (V, E')$ be the final stakeholder graph on which we compute three network measures.
A core hypothesis of the work is that \textit{who} identifies a risk matters as much as \textit{what} the risk is. We derive three primary rankings for prioritizing risks based on these network centrality measures:

\smallskip
\noindent \textbf{Degree Centrality Ranking.}
Risks are ranked by the sum of the Degree Centrality of the stakeholders who identified them:
$Score_{D}(R) = \sum_{s \in Contributors(R)} C_{deg}(s)$.
This favors risks identified by ``popular'' or most obvious stakeholders (e.g., Developers, Product Managers).

\smallskip
\noindent \textbf{Betweenness Centrality Ranking.}
Risks are ranked by the sum of the Betweenness Centrality of the stakeholders:
$Score_{B}(R) = \sum_{s \in Contributors(R)} C_{bet}(s)$. 
Betweenness centrality ($C_{bet}$) measures how often a node acts as a bridge along the shortest path between two other nodes. In our context, high-betweenness stakeholders are hypothesized to see risks that arise from stakeholders that bridge disconnected groups (e.g., ``Ethics \& Compliance Officer'' sitting between ``Regulatory Body'' and ``Developer'').

\smallskip
\noindent \textbf{Closeness Centrality Ranking.}
Risks are ranked by the sum of the Closeness Centrality of the stakeholders who identified them: 
%$$
$Score_{C}(R) = \sum_{s \in Contributors(R)} C_{clo}(s)$.
%$$
Closeness centrality ($C_{clo}$) captures how close a stakeholder is, on average, to all other stakeholders in the network, based on shortest path distances. This ranking strategy prioritizes risks raised by stakeholders who are well-positioned to rapidly integrate information across the network, favoring broadly informed perspectives rather than highly popular (degree) or structurally bridging (betweenness) roles.

In summary, these three ranking strategies reflect distinct theories of influence in collective risk assessment: degree centrality emphasizes visibility and consensus, betweenness centrality highlights cross-role mediation and systemic insight, and closeness centrality prioritizes globally informed perspectives.

Next, we empirically evaluate our proposed framework on an AI chatbot companion use case. \rev{We selected this specific use case because it represents a high-stakes, socio-emotional application characterized by wide demographic use and timely regulatory attention.}
%While focusing here allows us to rigorously evaluate complex, latent systemic risks, assessing the generalization of our framework to more situated, specialized use cases remains an important direction for future work. We evaluate our framework}
We evaluate our framework in two ways. First, we evaluate our framework with blank-slate of risks by AI practitioners (Section \ref{sec:eval1}). Second, we directly test our framework’s effectiveness in supporting human-led ideation in Participatory Risk Assessments through a controlled study using the Futures Wheel technique \cite{glenn_futures_wheel} (Section \ref{sec:futureswheel}), in which non-Western young chatbot users who started from risks generated by our framework are compared to those starting from practitioners-generated risks in our first evaluation (i.e., by AI practitioners).

\section{Evaluation of Our Framework for Generating Risks}
\label{sec:eval1}

\revision{To validate the efficacy of our framework, we evaluate whether networked simulation improves risk generation compared to flat baselines without network structures, using a dataset of practitioners-generated risks as a benchmark.}
%Section \ref{sec:eval1_workshop} details the collection of our practitioners-generated dataset through risk generation workshops. Section \ref{sec:rubric} defines the multi-dimensional rubric used by expert annotators to assess risk quality. Section \ref{sec:baselines} describes the single-LLM and multi-agent baselines against which we compare our framework. Finally, Section \ref{sec:results} presents the quantitative findings.}

\subsection{Establishing Brainstorming Workshops for Generating Practitioners-based Risks}
\label{sec:eval1_workshop}
\revision{To evaluate our framework, we drew on a dataset from 8 human-based brainstorming workshops focused on identifying risks associated with AI chatbot companions \cite{govers2026and}. Rather than representing an idealized participatory engagement, these workshops were designed to mimic the typical, early-stage internal risk workshops commonly conducted within public institutions and private companies. We recruited 45 AI practitioners across 3 locations via mailing lists, online ads, and EventBrite. All participants reported active personal use of AI chatbot companions, grounding their risk identification in everyday AI chatbot interaction. Those workshops utilized blank slate brainstorming (i.e., with no risks provided), each lasting 1 hour and 45 minutes with 4 to 7 participants. The study received institutional ethics approval and participants were compensated above minimum wage. } 
\revision{Participants (detailed demographics shown 
%in Table \ref{tab:chatbot_participants} 
in Appendix) were relatively young (majority aged 25–34), highly educated and all held university degrees (51\% Master’s, 20\% PhD). 
%Although approximately half of the cohort consisted of university students and researchers (many conducting research around topics related to AI), 
%The sample was neither homogeneous nor strictly academic: 
Many participants came from institutions and companies (37\% were from the company, 43\% researchers, and 20\% students). To capture a rich socio-technical perspective, the team specifically included: (1) \emph{AI professionals}, comprising software developers, NLP experts, AI monitoring/auditing experts, and designers, who contributed insights into technical and operational failure modes; (2) \emph{Applied AI researchers}, who developed AI algorithms in a company and evaluated their limitations; (3) \emph{Researchers in Responsible AI and psychology}, who specialized in the socio-emotional and downstream impacts of companions; and (4) \emph{Civil society representatives}, who advocated for protection of vulnerable groups.}
%broader societal safeguards and the protection of vulnerable groups.} 
%
%\revision{Crucially, the workshops fostered interdisciplinary ideation. While 60\% of participants possessed backgrounds in Computer Science and AI, the remaining participants mainly came from Health \& Life Sciences (13.3\%), Humanities \& Ethics (6.7\%), Business (6.7\%), and Physical Sciences \& Engineering (4.4\%). }
\revision{Self-reported AI expertise was high, with 65\% of participants reporting knowledgeable or expert levels. All workshops were balanced to include company workers, researchers, and students. Across all sessions, participants generated 244 distinct risks, which form the basis of our evaluation.}

\subsection{Defining the Multi-Dimensional Rubric} %Evaluation 
\label{sec:rubric}

We adopted a multi-dimensional rubric (Table~\ref{tab:rubric}) to evaluate the quality of practitioners-generated risks for chatbot companion within those brainstorming workshops. 
\revision{Synthesizing criteria from the NIST AI Risk Management Framework and other prior AI risk assessment frameworks \cite{ai2023artificial, plausibility, kieslich2025scenario, novelty, engagement_1, engagement_2}, the rubric assesses plausibility, probability, severity, uniqueness, novelty, and usability.} 
\rev{To explicitly distinguish between evaluation constructs, our rubric separates Plausibility (theoretical feasibility), Likelihood (realistic probability of occurrence in deployment), and Uniqueness (specificity across domains), a correlation matrix provided in Appendix 
%(Figure \ref{app:prolific}) 
demonstrates that these dimensions capture distinct evaluative signals rather than redundant measures.}
Five external AI-domain experts recruited via Prolific rated the risks using Likert scales, following attention and comprehension screening. \rev{They are AI ethicists who frequently develop assessments for AI.} 
For consistency and ease of comparisons, all the ratings were normalized to five-point Likert scale. 
These expert annotations provide a quantitative basis for comparing risk quality across brainstorming methods and with system-generated outputs.

\begin{table}[t]
%\centering
\footnotesize
%\scriptsize
%\begin{tabular}{lp{6.3cm}}
\begin{tabular}{@{}p{1.3cm}p{6.6cm}@{}}
\toprule
\textbf{Dimension} & \textbf{Definition} \\
\midrule

\textbf{Plausibility} 
& Assesses whether the described risk could realistically arise in the given AI use context (1--5 Likert \cite{plausibility}). \\

\textbf{Likelihood} 
& Estimates the likelihood that a risk will occur (1--7 Likert, based on NIST AI risk assessment measures~\cite{ai2023artificial} and foresight \cite{frohling2026agent}). \\

\textbf{Uniqueness} 
& Evaluates whether the risk is specific to the particular AI use case, rather than broadly applicable across AI systems or domains (1--3 Likert~\cite{kieslich2025scenario}). \\

\textbf{Novelty} 
& Rates the degree of creativity and non-obviousness of the risk, ranging from mundane or checklist-like to imaginative and original (1--3 Likert~\cite{novelty, frohling2026agent}, ). \\

\textbf{Usability} 
& Measures how clear, understandable, and practically engaging the risk (1--5 Likert~\cite{engagement_1, engagement_2}). \\

\textbf{Severity} 
& Assesses the seriousness of potential harm and its downstream social, economic, or institutional consequences (1--5 Likert~\cite{ai2023artificial, frohling2026agent}). \\

%Assesses the seriousness of potential harm if a risk were to materialise, considering downstream social, economic, or institutional consequences (1--5 Likert~\cite{ai2023artificial, frohling2026agent}). \\

\bottomrule
\end{tabular}
\caption{Evaluation rubric used for annotating AI risks. Each dimension corresponds to explicit annotation questions posed to evaluators, and is grounded in prior literature.} 
\label{tab:rubric}
\end{table}

\begin{figure*}[t]
    \centering
    \includegraphics[width=0.9\linewidth]{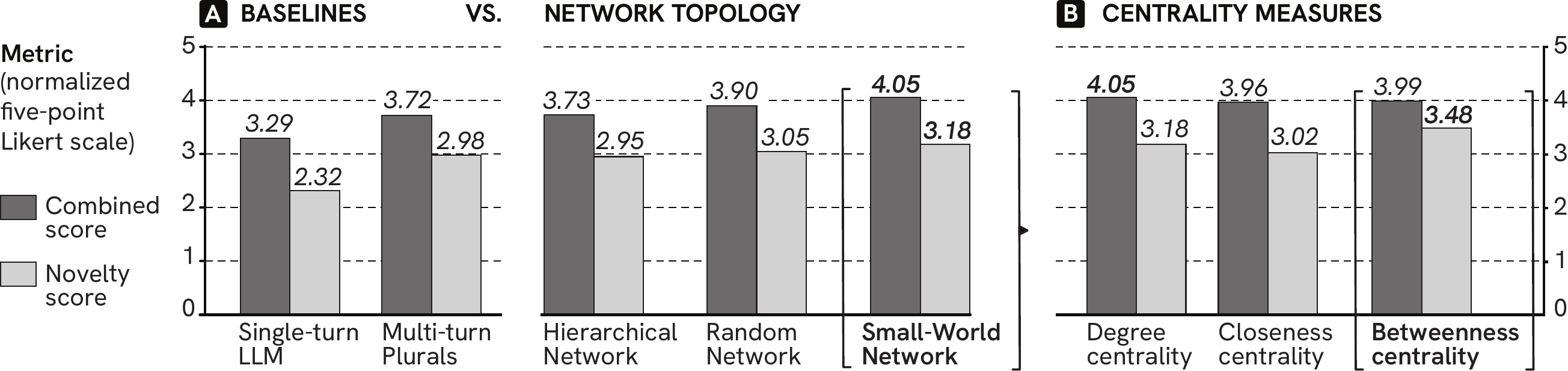}
    %\caption{Our analysis utilizes two primary metrics: ``Combined Score'' aggregates Plausibility, Likelihood, Uniqueness, Usability and Severity while ``Novelty'' solely reflects risk diversity. All measures were normalized to the five-point Likert scale, and the top performing systems are bolded. \textbf{A}: Comparison of Network Topologies (using Snowballing Typical Roles and Degree Centrality): Small-World networks outperform both structured hierarchies and unstructured random networks. They significantly outperform the flat Baselines, demonstrating the value of structured interaction.
    %\textbf{B}: Impact of Centrality Measures on Risk Quality (Small-World Network + Snowballing Typical Roles): Degree Centrality yields the highest overall quality (Combined Score) by capturing consensus; Betweenness Centrality is most effective at surfacing Novel risks by prioritizing risks from bridging stakeholders.} 
    \caption{Our analysis utilizes two metrics: ``Combined Score'' aggregates Plausibility, Likelihood, Uniqueness, Usability and Severity while ``Novelty'' reflects risk diversity. All measures were normalized to the five-point Likert scale, and the top performing systems are bolded. \textbf{A}: Comparison of Network Topologies (using Snowballing Typical Roles and Degree Centrality): Small-World networks outperform both structured hierarchies and unstructured random networks. They significantly outperform the Baselines, demonstrating the value of structured interaction.
    \textbf{B}: Impact of Centrality Measures on Risk Quality (Small-World Network + Snowballing Typical Roles): Degree Centrality yields the highest quality (Combined Score) by capturing consensus; Betweenness Centrality is most effective at surfacing Novel risks by prioritizing risks from bridging stakeholders.}
    \label{fig:network_comparison}
\end{figure*}

\subsection{Establishing Single-/Multi-Agent Baselines}
\label{sec:baselines}

We compare our approach against two baselines commonly used in AI risk assessment:

\smallskip
\noindent \textbf{Single-Turn LLM Expert Prompting}: A single LLM is prompted to generate and rank risks from an expert perspective, conditioned on a description of the AI use case system and a list of stakeholders. This single LLM approach reflects current practice in automated risk generation tools, such as Farsight \cite{wang2024farsight}, ExploreGen \cite{herdel_et_al_exploregen25}, and AHA! \cite{buçinca_aha_generating_ai_risks23}.

\smallskip
\noindent \textbf{Multi-Turn Plurals without Network Awareness}: We use standard Plurals framework \cite{ashkinaze2025plurals} to simulate multiple in-silico stakeholder agents interacting in a flat ensemble structure, without explicit network structure (topology) or network-based ranking . This baseline isolates the contribution of network structure and measures, and represents current agentic LLM brainstorming method.

To evaluate those two baselines and our framework, we mapped all system-generated risks to practitioners-generated risks using a two-stage process combining embedding-based similarity and manual validation. Each of the 244 practitioners-curated risks was embedded and matched to the five most similar system outputs, producing over 1,000 candidate pairs. We manually validated semantic matches, allowing one-to-many mappings.
In total, 108 human risks (44\%) matched at least one system-generated risk. These matched risks were used for our evaluation.
%\footnote{In addition to the matched human annotation we used for our evaluation, we also performed an LLM-based evaluation (i.e., LLM as a judge \cite{zheng2023judging}) on the full set of system-generated risks, including those risks not matched in the human workshops, using the same rubric dimensions (Table \ref{tab:rubric}). The results were consistently similar to the findings presented in the Results section.}

\subsection{Results: Network Prioritization Drives Novelty}
\label{sec:results}

We evaluate our framework against static and unstructured baselines using two metrics: a \textit{Combined Score} (an aggregate of plausibility, likelihood, uniqueness, usability, and severity), and a \textit{Novelty} score (specifically isolating risk diversity and non-obviousness).  All scores are normalized to a five-point Likert scale. \rev{We explicitly report \textit{Novelty} score to capture non-obvious risks, often disproportionately affecting minority or marginalized populations, that are systematically obscured by consensus-driven aggregates.}

By moving beyond flat lists of stakeholders, we show that network topology, centrality measures, and discovery strategies fundamentally shape the quality of generated risks (evaluation results on all system variants are provided in Appendix  with a detailed discussion of the discovery strategies). We highlight our key findings below.

\begin{figure*}[t]
    \centering
    \includegraphics[width=1.0\linewidth]{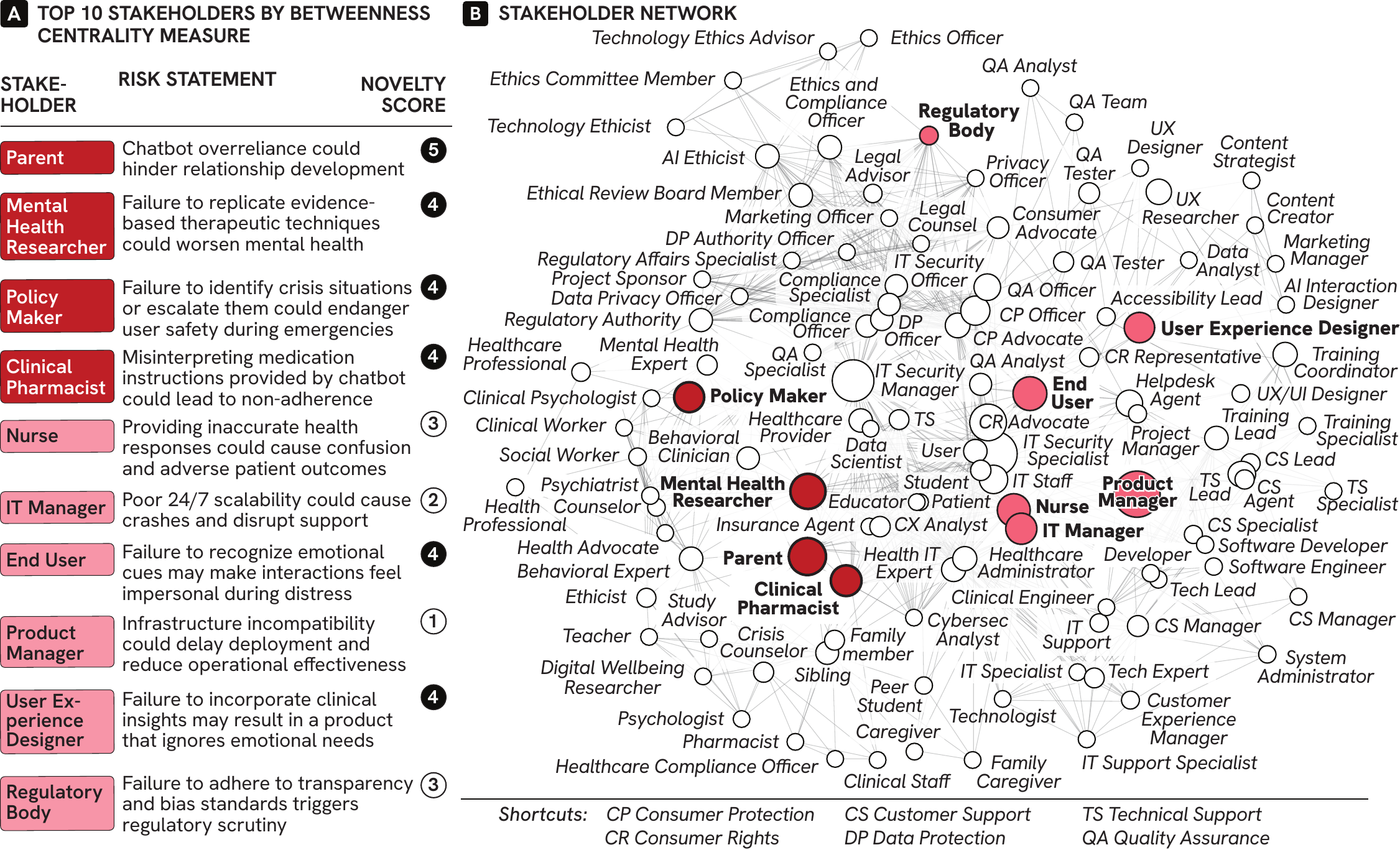}
    \caption{\textbf{A:} Risks and their novelty scores generated by the top 10 stakeholders ranked by Betweenness Centrality. \textbf{B:} The stakeholder network constructed through snowballing, within which in-silico LLM agents are connected and communicate to brainstorm risks. Many stakeholders who bridge diverse domains with top Betweenness Centrality such as Parent, Mental Health Researchers, Policy Makers and Clinical Pharmacist are the primary identifiers of highly novel risks (scores 4–5) such as psychological dependency and user safety during crisis.} %\todo{change top 1 example}}
    \label{fig:roles-dag}
\end{figure*}

\smallskip
\noindent \textbf{Small-World Topologies Optimize Ideation Efficiency.} Network science literature \cite{watts1998collective} posits that ``small-world'' networks, characterized by high local clustering and short average path lengths, are uniquely efficient at information diffusion. We find that this structural advantage holds for simulated AI risk ideation, significantly outperforming unstructured interactions.
Figure \ref{fig:network_comparison}A compares three network topologies using the same strategy for creating the set of stakeholders: 

\noindent \emph{Small-World network} achieves the best balance (Combined 4.05, Novelty 3.18). The high local clusters allows distinct stakeholder groups (e.g., a ``Security Cluster'' \emph{vs.} a ``Patient Advocacy Cluster'') to develop coherent, deep arguments locally before propagating them.

\noindent \emph{Random network} performs competitively on the Combined Score but suffer on Novelty (3.05). We found stakeholders lack the local context to refine specific critiques, leading to more generic outputs.

\noindent \emph{Hierarchical network} performs poorly on Novelty (2.95). Designed to mimic corporate reporting lines, these structures appear to introduce information filtering. ``Middle manager'' agents may inadvertently smooth over radical 
signals from leaf nodes (end-users) before they reach the central nodes, effectively suppressing diverse risk identification.

The structural impact is evident when comparing Small-World results to the \textit{Single-turn LLM} (Combined 3.29) and \textit{Multi-turn Plurals} (Combined 3.72). The Single-turn baseline lacks the iterative refinement of ideas, while the Multi-turn baseline lacks the community structure to support specialized discourse. The Small-World topology provides a +0.76 
improvement in Combined Score over the standard LLM baseline, 
demonstrating that \textit{how} agents are connected is as critical as the agents themselves.

\noindent \textbf{Novelty is Driven by Bridging Stakeholders (Using Betweenness Centrality).}
A core challenge in AI auditing is moving beyond ``known knowns'' (high-consensus risks) to uncover ``unknown unknowns''. Standard methods often converge on high-likelihood but generic risks \cite{devos2022toward, lancaster2024s}. We find that the choice of network centrality measure used to prioritize risks alters the nature of the output, revealing a trade-off between consensus and discovery (Figure \ref{fig:network_comparison}B)
Ranking risks by \textit{Degree Centrality} (popularity) yields the highest Combined Scores (4.05). However, qualitative analysis suggests these risks often represent consensus views, broadly recognized issues like data privacy leaks or standard bias. While these are valid, they are often already known to development teams.
In contrast, prioritizing risks via \textit{Betweenness Centrality} results in the highest \textit{Novelty} scores (3.48). Betweenness centrality identifies stakeholders who act as bridges between otherwise disconnected sub-communities.
Unlike high-degree nodes that reinforce dominant narratives, bridging nodes are uniquely positioned to translate context across epistemic boundaries.

To illustrate this mechanism, we analyzed the risks generated by the top 10 stakeholders ranked by Betweenness Centrality (Figure~\ref{fig:roles-dag}). Some of those top 10 stakeholders also overlap with those in top 10 of Degree Centrality measures such as IT Manager and Product Manger. The data reveals a stark divergence in Novelty depending on the \textit{type} of domain the stakeholder bridges.
Stakeholders who also has high Degree Centrality tended to identify valid but low-novelty risks. For example, the \emph{Product Manager} identified the risk of \textit{``Incompatibility with existing infrastructure''} (Novelty=1), and the \emph{IT Manager} identified scalability risks regarding \textit{``24/7 volume causing crashes''} (Novelty=2). Their outputs reflect standard operational concerns common in software deployment.
In contrast, stakeholders who are mainly bridging sub-communities with extremely high Betweenness Centrality but relatively low Degree Centrality
provided the highest novelty risks:

    \noindent The \emph{Parent} (Novelty 5), acting as a bridge between the domestic sphere and the educational/technical system, identified the risk of \textit{``Developing an unhealthy over-reliance on the chatbot''}, hindering real-life social development. This is a profound socio-psychological harm that purely technical agents missed, although this harm is documented in prior literature (e.g., loneliness-through-technology \cite{wilson2018love}).
    %\footnote{\rev{We acknowledge that ``novelty'' in this context denotes specificity and non-obviousness in this deployment, not an absolute absence from prior literature (e.g., loneliness-through-technology is a well-known harm \cite{wilson2018love}).}}
    
    %\noindent The \emph{Mental Health Researcher} (Novelty 4) and \emph{Clinical Pharmacist} (Novelty 4) bridged clinical standards with AI interaction. They identified nuanced risks such as the \textit{``Inability to effectively replicate evidence-based therapeutic techniques''}, and the danger of \textit{``Misinterpreting medication instructions''}, respectively. 

    \noindent The \emph{Mental Health Researcher} and \emph{Clinical Pharmacist} (both Novelty 4) bridged clinical standards and AI interaction, identifying risks such as \textit{``Inability to effectively replicate evidence-based therapeutic techniques''} and \textit{``Misinterpreting medication instructions''}, respectively.
    
    \noindent The \emph{Policy Maker} (Novelty 4), bridging public safety and user interaction, highlighted the systemic failure to \textit{``identify crisis situations or escalate them''}, a critical safety gap. % during emergencies.

This analysis confirms that novelty is not randomly distributed. It is structurally concentrated in stakeholders who navigate the boundaries between distinct communities. By prioritizing these bridging nodes via Betweenness Centrality, the system identifies high-impact, non-obvious risks that are invisible to the commonly identified risks.

\section{\revision{Evaluation of Our Framework Supporting Participatory AI Risk Assessment}}
\label{sec:futureswheel}

\revision{In this section, we focus on testing whether our framework actually helps a human-led ideation session.} 
\revAIES{The goal of this evaluation was to test whether seeding a participatory session with network-generated risks leads participants to identify more risks, particularly systemic and downstream risks, than seeding the session with practitioner-generated risks.} 
%Section \ref{sec:futureswheel_setup} outlines the experimental design, detailing the Futures Wheel methodology, participant demographics, and the control versus treatment conditions, while Section \ref{sec:futureswheel_results} presents the main findings.

\subsection{\revision{Participatory AI Risk Assessment with Futures Wheel}}
\label{sec:futureswheel_setup}

\revision{We conducted a user study using the Futures Wheel method \cite{glenn_futures_wheel}. It is a structured brainstorming technique for identifying first-, second-, and third-order consequences of a central topic, enabling participants to explore cascading and interconnected effects. Starting from an AI chatbot companion, participants iteratively answer the question ``If this occurs, what happens next?'', generating primary consequences in the first layer, followed by secondary and tertiary consequences in the second and third layers. This process encourages participants to move beyond immediate risks toward more systemic and long-term impacts.}

\revision{We selected the Futures Wheel as our method of participatory AI risk assessment for three reasons. First, it offers a simple, structured format for team ideation that enables participants to generate a large number of risks without requiring domain expertise \cite{glenn_futures_wheel}. Second, it pushes participants to think beyond obvious risks for AI chatbots by exploring their second- and third-order consequences \cite{bengston2016futures}. Third, it has been successfully used in large-scale participatory AI risk assessment studies (with over 250 participants across six Global Majority and Minority countries \cite{GlobalAIDialogues2025}), making it appropriate for our intended participants.}

\smallskip
\noindent \textbf{Participant Demographics and Recruitment.}
\revision{We are particularly interested in testing how our framework supports risk ideation among a crucial but under-represented demographic in AI risk literature: young, frequent chatbot users from predominantly non-Western (non-US, non-European) backgrounds \cite{png2022tensions, birhane2022power}. Young users in the Global South and non-Western contexts are major consumers of AI companions \cite{scherr2025explaining}, yet their perspectives are rarely centered in participatory risk assessments \cite{png2022tensions, birhane2022power}. }
% .
%
\revision{We recruited 11 teams of participants, comprising a total of 40 individuals (with three to four people per team). Participants generally reported high AI chatbot usage: with 20 participants (50\%) using AI chatbots multiple times per day, 6 (15\%) once daily, 11 (27.5\%) 2–6 times per week, and 3 (7.5\%) once per week. Additionally, 13 participants (32.5\%) reported having experienced technology-related harms, citing issues such as misinformation, privacy loss, social media addiction, surveillance, and over-reliance on technology. Ethnically, the cohort was diverse: 15 participants (37.5\%) identified as Middle Eastern or North African, 10 (25\%) as Asian, 8 (20\%) as White, 4 (10\%) as Black, 1 (2.5\%) as Mixed, 1 (2.5\%) as Central Asian, and 1 (2.5\%) preferred not to say.} %In terms of education, 29 participants (72.5\%) held a Bachelor’s degree and 11 (27.5\%) held a Master’s degree.} 

\begin{figure}[t]
    \centering
    \includegraphics[width=\linewidth]{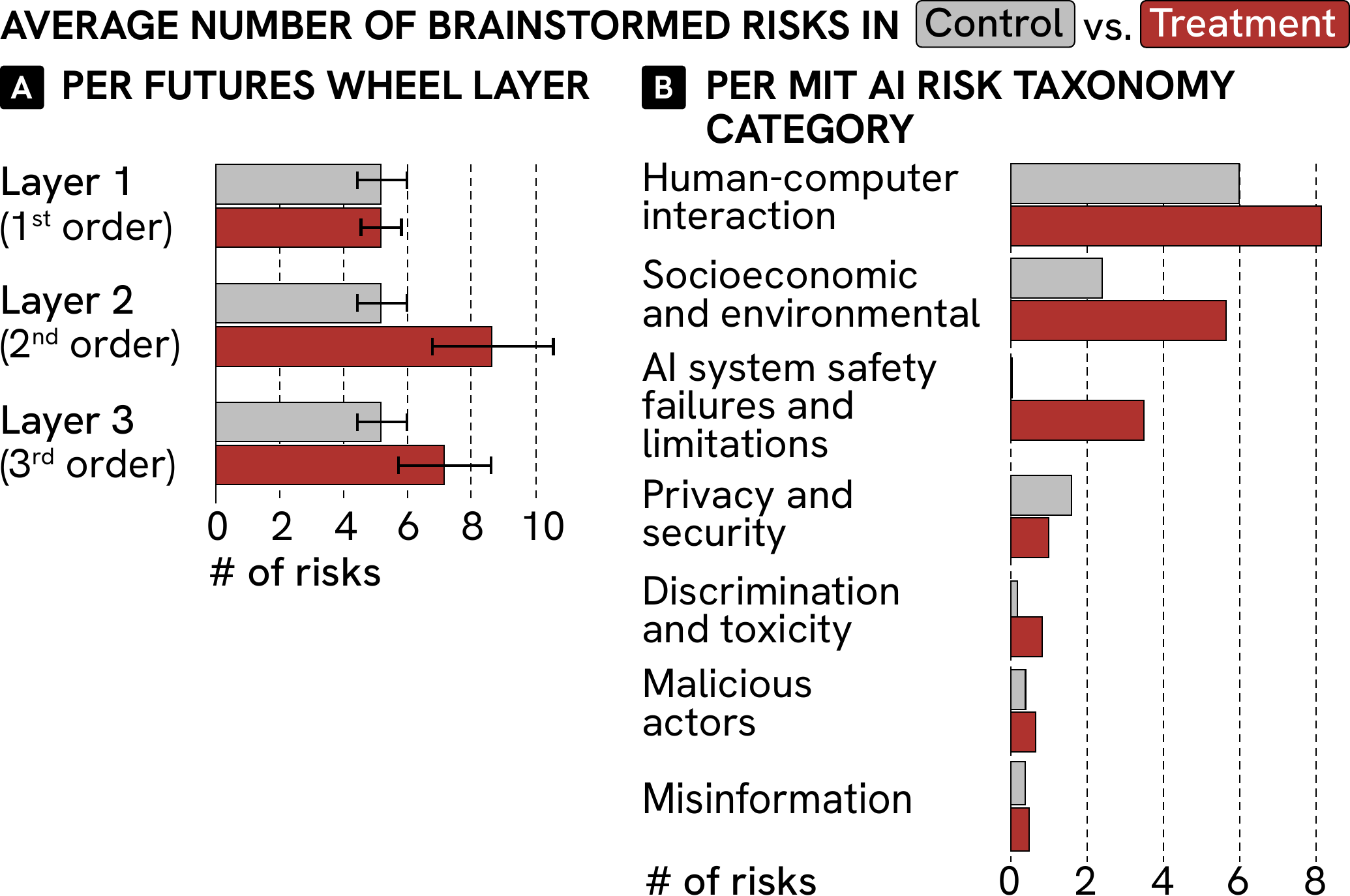}
    %\caption{\revision{Results from the Futures Wheel participatory ideation study comparing teams seeded with practitioners-generated risks (Control) versus our networked simulation generated risks (Treatment): \textbf{A:} Average number of risks brainstormed per team across the three layers of the Futures Wheel.  \textbf{B:} Average number of risks per team mapped to the MIT AI Risk taxonomy \cite{slattery2024ai}. The treatment teams identified significantly more complex second- and third-order risks, moving beyond obvious surface-level issues, and uncovered a much wider range of socio-technical harms, including human-computer interaction and long-term socioeconomic and environmental risks, extending well beyond conventional privacy and security risks.}}
    \caption{\textbf{Futures Wheel participatory ideation results.} Teams seeded with our networked simulation generated risks (Treatment) identified more second- and third-order risks (A) than teams seeded with practitioner-generated risks (Control). They also surfaced a broader range of socio-technical harms, particularly human–computer interaction and socioeconomic and environmental risks (B).}
    \label{fig:ideation}
\end{figure}

\smallskip
\noindent \textbf{\revision{Experimental Conditions: Practitioners-Seeded (Control) \emph{vs.} Framework-Seeded Ideation (Treatment).}}
\revision{The teams were randomly assigned to two conditions: \emph{Control} teams: five teams were provided with a report containing a set of risks brainstormed by AI practitioners in our first evaluation,
%(see Figure \ref{fig:report_baseline} in Appendix \ref{appendix:reports})
 i.e., blank-slate risk brainstorming workshops as described in Section \ref{sec:eval1_workshop}; \emph{Treatment} teams: six teams were instead provided with a report containing the same number of risks 
 %(see Figure \ref{fig:report_treatment} in Appendix \ref{appendix:reports}) 
 but generated by our framework with a small-world network topology (Section \ref{sec:networkagent}). All the risks were randomly sampled from the set of top prioritized risks from each approach to ensure coverage of the wide spectrum. All teams then conducted the participatory AI risk assessment ideation exercise, brainstorming risks across three layers of consequences in the Futures Wheel.}

\subsection{\revision{Results: Our Framework Identifies More and Deeper Systemic Risks}}
\label{sec:futureswheel_results}

\revision{Overall, the treatment teams generated a substantially higher volume of risks compared to the control teams. On average, treatment teams brainstormed 21.1 risks per team, compared with 15.6 risks among control teams. Crucially, as shown in Figure \ref{fig:ideation}A, this increase in volume was entirely driven by a greater capacity to uncover higher-order, systemic consequences, which tend to be harder to envision \cite{friedman2012envisioning}. 
Both conditions yielded an identical average of 5.2 risks at Level 1 (immediate consequences). However, the treatment group significantly outperformed the control group in subsequent layers, identifying an average of 8.7 risks at Level 2 (compared to 5.2 in the control) and 7.2 risks at Level 3 (compared to 5.2 in the control). This demonstrates that our framework effectively prompts human participants to think deeper and anticipate cascading effects rather than exhausting their effort on obvious primary outcomes.}

\revision{To understand which risks were identified, we classified the brainstormed risks using the MIT AI Risk taxonomy \cite{slattery2024ai} (Figure \ref{fig:ideation}B). We found that the treatment teams generated more risks across almost all categories. Most notably, they surfaced substantially more risks in complex, downstream areas such as Human-Computer Interaction (8.2 \emph{vs.} 6.0), Socioeconomic \& Environmental harms (5.6 \emph{vs.} 2.4), and AI System Safety Failures \& Limitations (3.5 \emph{vs.} 0.0). Interestingly, the control teams only produced more risks in the ``Privacy \& Security'' category (1.5 \emph{vs.} 1.0). This 
%aligns with our earlier findings 
demonstrates practitioners-seeded baselines often anchor human ideation on well-known, conventional risks, whereas the networked stakeholder simulation encourages the exploration of non-obvious, systemic harms.}

\revision{Ultimately, these results indicate that utilizing our framework to support the ideation stage of participatory AI risk assessments is significantly more effective than relying on a baseline of practitioner-generated risks. It not only increases the total number of identified risks, but specifically enhances human capacity to discover the higher-order socio-technical harms that are typically the hardest to spot.}

\section{Conclusion and Discussion}

We presented a framework that transforms AI risk assessment %from a static checkbox exercise 
into a dynamic, multi-perspective exploration. By generating stakeholder networks and utilizing social network analysis measures like Betweenness Centrality, we demonstrated a capability to uncover high-impact risks that traditional methods miss.
\revision{Our findings suggest that AI risk assessment benefits not only from who is involved, but from how stakeholder perspectives, even in-silico, are structured, connected, and weighted. By treating risk identification as a network-mediated ideation, we provide practitioners with a tool to help with a human-led ideation session in the participatory AI risk assessments.}
\smallskip

\noindent \textbf{Implications.} This work has three key implications:

\smallskip
\noindent \emph{From flat independent assessments to network-mediated ideation.} Current AI risk assessment practices, whether human-led brainstorming workshops or LLM-generated risk assessments, implicitly assume that risks are enumerable, independent, and equally visible to all evaluators. Risks emerge at the intersection of stakeholder roles (e.g., between legal compliance and user vulnerability, or between product incentives and mental health outcomes) are systematically underrepresented by flat generation methods. By modeling stakeholder interaction explicitly, our approach shifts the focus from coverage of known risk categories to discovery of emergent, relational harms.

\smallskip
\noindent \emph{\revision{Scalable pluralism and concrete usage pathways.}} \revision{Our framework is explicitly designed as a complementary pre-assessment tool, not a substitute for authentic participation in participatory risk assessment. Risk management professionals can integrate this tool into early design sprints to conduct structured gap analyses against regulatory taxonomies (e.g., the EU AI Act). By running the simulation prior to a human workshop, organizers can use the pre-generated materials, specifically areas of simulated disagreement and heavily populated systemic risks, as prompts for human ideation as we have done in our second evaluation. This prevents human participants from starting with a ``blank slate'' and focuses their valuable time contextualizing, and deepening complex risks.} \revAIES{The tool allows organizers to explicitly balance AI practitioners' operational insights with the broader, theoretically informed views of researchers.}
\rev{If the simulation reveals that a specific bridging stakeholder (e.g., a ``Mental Health Advocate'') is central to identifying high-severity risks, organizers are provided an evidence-based justification to invest resources into recruiting humans holding those lived experiences for subsequent participatory stages.}
\revision{We explicitly caution against using our tool to bypass authentic inclusion with actual human participation for the full risk assessment.} \rev{Most importantly, outputs from this tool must not be submitted to auditors or regulators as a substitute for authentic stakeholder evidence or testimony.}

\noindent \emph{Prioritization as a normative choice.} Our comparison of centrality network based ranking strategies highlights that risk prioritization is not value-neutral. Degree-based ranking favors widely recognized, consensus risks; Betweenness-based ranking elevates boundary-spanning concerns. This trade-off mirrors real-world governance tensions between addressing ``known knowns'' and anticipating low-frequency but high-impact harms (potentially to marginalized stakeholders). Making these prioritization logics explicit allows practitioners and regulators to align risk assessment outputs with their normative goals.

\smallskip
\noindent \textbf{Limitations.}  
Our work has five main limitations.

\revAIES{Firstly, our framework relies on LLM-simulated stakeholders to model structured information flow among network-conditioned agents, rather than authentic stakeholder deliberation \cite{habermas1981theory}. Because these in-silico representations cannot fully capture the lived experience, contextual knowledge, or normative authority of real affected communities, certain harms, particularly those rooted in cultural, emotional, or marginalized perspectives, may be misrepresented or missed.}

\rev{Secondly, while our Futures Wheel study grounded the framework in a human-seeded task, it involved only one stakeholder type (young, non-Western users). Crucial domain experts and vulnerable groups, such as mental health clinicians and social workers, were not included. Systematically benchmarking against workshops involving both highly vulnerable stakeholders and these specialized domain experts remains a critical direction for future work.}

\revAIES{Thirdly, the quality and diversity of generated risks depend on the underlying language model and the architectural choices made by the system designers. This introduces a structural paradox: while the tool is intended to mitigate the narrow biases of traditional human brainstorming, we (as researchers) defined the network topologies and prompt structures, which might inevitably encode our own biases into the system. Compounding this issue, because LLMs often encode representational biases from historically skewed training data \cite{bai2025explicitly, gupta2023bias, shujaa2025llms}, simulated stakeholders run the risk of producing algorithmic caricatures rather than authentic reflections of marginalized communities \cite{vecchione2021algorithmic}.}
Procedurally, it is imperative that our framework is strictly utilized as a pre-assessment tool to seed human ideation, not as a substitute for authentic human participation. A vital direction for future work is to systematically benchmark these stereotyping effects and validate our simulations directly against workshops involving highly vulnerable, directly affected stakeholders to ensure the tool safely bridges, rather than widens, representational gaps.

Fourthly, the constructed stakeholder networks encode simplifying assumptions about influence flow (e.g., acyclicity), which may not fully reflect the reciprocal and evolving nature of real-world stakeholder dynamics.

Finally, our evaluation focuses on a single use case (AI chatbot companions), and further studies are needed to assess the generalizability of the approach across domains with different stakeholder structures and risk profiles.

\section*{Ethics Statement}
All human participants, including the AI practitioners in the baseline study and the annotators recruited via Prolific, provided informed consent prior to participation. No personally identifiable information was collected, and all data was anonymized prior to analysis. Participants in the baseline and Futures Wheel evaluation phases were compensated exceeding the local minimum wage. For the expert verification tasks on Prolific, stringent screening criteria were applied to ensure domain expertise.

\bibliography{main}

\appendix
\onecolumn
\clearpage

\section{Appendix}

\subsection{Implementations}

To operationalize the complex interactions required for our stakeholder simulation, we adopt and extend the Plurals framework proposed by Ashkinaze et al. \cite{ashkinaze2025plurals}. Plurals is a modular system designed to steer Large Language Models via simulated recursive social deliberation. It abstracts the chaotic process of multi-agent conversation into three governable primitives, which we map to our risk assessment task:

\emph{Agents (The Stakeholders):} In Plurals, agents are LLM instances initialized with specific ``system\_prompts'' and ``personas''. We map each node $s$ in our stakeholder graph to a Plurals Agent, injecting the specific role description (e.g., ``software engineer'') into their initialization state.
    
\emph{Structures (The Topology):} The framework manages information flow through defined topologies (e.g., Ensembles, Debates, Chains). While the original framework focuses on static interaction patterns, we extend this by dynamically feeding our constructed network structures  (Hierarchical, Small-World, Random) into the Plurals graph executor. This ensures that an Agent only receives context from its specific neighbors defined in our network generation phase (i.e., local connected network), rather than a global context window.
    
\emph{Moderators (The Synthesizers):} These are specialized agents tasked with processing the output of a deliberation round. We utilize Moderators not to force consensus, as is common in political deliberation tasks, but to perform \textit{deduplication} and \textit{formatting} of risks between rounds, preventing the ``echo chamber'' effect where agents merely repeat their neighbors' inputs.

Our implementation utilizes the Plurals Library (\url{https://josh-ashkinaze.github.io/plurals/}) to handle the asynchronous orchestration of these agents, while our contribution lies in the dynamic generation of the \textit{Structure} (via snowballing) and the post-hoc analysis of the graph using network measures for prioritizing risks.
\revision{To address the known representational biases in LLMs, we implemented targeted prompt diversification. Agent prompts explicitly instructed the LLMs to adopt an exploratory and uncertainty-aware framing (e.g., ``Identify potential socio-technical risks that *might* uniquely affect this stakeholder group''). Moderators were prompted to synthesize without erasing minority dissenting views, counteracting the LLM tendency to collapse into consensus stereotypes.}

\subsection{Supplementary Experimental Results}
\label{app:exp}

Table~\ref{tab:results_measures_full} summarizes the performance of all system variants across network structures, stakeholder strategies, and ranking measures based on two metrics: (1)\emph{``Combined Score''} that averages Plausibility, Likelihood, Uniqueness, Engagement and Severity, and (2) \emph{``Novelty''} that solely reflects risk diversity. All measures were normalized to five-point Likert scale.

\begin{table*}[t]
\small
\centering
\caption{Performance of AI risk assessment system variants, organized by network variant (structure + measure). ``Combined Score'' aggregates Plausibility, Likelihood, Uniqueness, Engagement and Severity while ``Novelty'' solely reflects risk diversity. All measures were normalized to the five-point Likert scale, and the top-two performing systems are bolded. We observe that performance improves systematically with (i) small-world interaction structures, (ii) dynamic stakeholder expansion, and (iii) network-aware prioritization. Degree centrality yields the highest combined risk scores, whereas betweenness centrality surfaces the most novel and diverse risks.} %\todo{Break down further on individual scores.}}
\label{tab:results_measures_full}
\begin{tabular}{llcc}
\toprule
\textbf{Network Variant (Structure + Measure)} &
\textbf{Roles (Stakeholder Strategy)} &
\textbf{Combined Score} &
\textbf{Novelty} \\
\midrule

\multicolumn{4}{l}{\textit{Baselines}} \\
None 
& Single-turn LLM
& 3.29 
& 2.32 \\

Implicit DAG 
& Multi-turn Plurals
& 3.72 
& 2.98 \\

\midrule
\multicolumn{4}{l}{\textit{Hierarchical Network }} \\

Hierarchy 
& Fixed Typical Roles (5 roles) 
& 3.15 
& 2.21 \\

Hierarchy 
& Fixed Comprehensive Roles 
& 3.58 
& 2.70 \\

Hierarchy 
& Snowballing Typical Roles 
& 3.73 
& 2.95 \\

Hierarchy 
& Snowballing Comprehensive Roles 
& 3.77 
& 3.00 \\

\midrule
\multicolumn{4}{l}{\textit{Small-World Network + Degree Centrality}} \\

Small-world + Degree 
& Fixed Typical Roles (5 roles) 
& 3.26 
& 2.17 \\

Small-world + Degree 
& Fixed Comprehensive Roles 
& 3.78 
& 2.95 \\

Small-world + Degree 
& Snowballing Typical Roles 
& \textbf{4.05} 
& 3.18 \\

Small-world + Degree 
& Snowballing Comprehensive Roles 
& 4.01 
& 3.22 \\

\midrule
\multicolumn{4}{l}{\textit{Small-World Network + Betweenness Centrality}} \\

Small-world + Betweenness 
& Fixed Typical Roles (5 roles) 
& 3.19 
& 2.36 \\

Small-world + Betweenness 
& Fixed Comprehensive Roles 
& 3.73 
& 3.12 \\

Small-world + Betweenness 
& Snowballing Typical Roles 
& \textbf{3.99} 
& \textbf{3.48} \\

Small-world + Betweenness 
& Snowballing Comprehensive Roles 
& 3.94 
& 3.42 \\

\midrule
\multicolumn{4}{l}{\textit{Small-World Network + Closeness Centrality}} \\

Small-world + Closeness 
& Fixed Typical Roles (5 roles) 
& 3.12 
& 2.32 \\

Small-world + Closeness 
& Fixed Comprehensive Roles 
& 3.67 
& 2.78 \\

Small-world + Closeness 
& Snowballing Typical Roles 
& 3.96 
& 3.02 \\

Small-world + Closeness 
& Snowballing Comprehensive Roles 
& 3.91
& 3.06 \\

\midrule
\multicolumn{4}{l}{\textit{Random Network + Degree Centrality}} \\

Random + Degree 
& Fixed Typical Roles (5 roles) 
& 3.14 
& 2.33 \\

Random + Degree 
& Fixed Comprehensive Roles 
& 3.58 
& 2.86 \\

Random + Degree 
& Snowballing Typical Roles 
& 3.90 
& 3.05 \\

Random + Degree 
& Snowballing Comprehensive Roles 
& 3.82 
& 3.10 \\

\midrule
\multicolumn{4}{l}{\textit{Random Network + Betweenness Centrality}} \\

Random + Betweenness 
& Fixed Typical Roles (5 roles) 
& 3.07 
& 2.38 \\

Random + Betweenness 
& Fixed Comprehensive Roles 
& 3.51 
& 3.02 \\

Random + Betweenness 
& Snowballing Typical Roles 
& 3.78 
& 3.22 \\

Random + Betweenness 
& Snowballing Comprehensive Roles 
& 3.75 
& 3.28 \\

\midrule
\multicolumn{4}{l}{\textit{Random Network + Closeness Centrality}} \\

Random + Closeness 
& Fixed Typical Roles (5 roles) 
& 3.06 
& 2.19 \\

Random + Closeness 
& Fixed Comprehensive Roles 
& 3.48 
& 2.72 \\

Random + Closeness 
& Snowballing Typical Roles 
& 3.76 
& 2.95 \\

Random + Closeness 
& Snowballing Comprehensive Roles 
& 3.68 
& 3.00 \\

\bottomrule
\end{tabular}
\end{table*}

\subsubsection{Generic Seeding Enables Broader Stakeholder Discovery}

%Finally, 
We examine the initialization strategy: is it better to manually curate a nuanced list of stakeholders, or to seed the system with a few generic roles and allow it to ``snowball'' (recursively discover) new ones? We focus this analysis using \textit{Betweenness Centrality} (Table~\ref{tab:seeding}), as this measure best captures the value of discovering bridging stakeholders.

We find that starting with a \textit{Typical} (generic) stakeholder set and utilizing snowballing is remarkably effective. As shown in Table~\ref{tab:seeding}, the ``Typical + Snowballing'' configuration achieves a higher Combined Score (3.99) than the ``Nuanced + Snowballing'' configuration (3.94), while maintaining a relatively high Novelty score (3.48 vs 3.42).
This is a counter-intuitive but significant finding. It suggests that providing a nuanced list \textit{ex ante} may induce path dependency, constraining the LLM to a pre-defined search space and limiting the ``organic'' discovery of bridging nodes. In contrast, generic seeding acts as a ``minimum viable prompt''. When the system is allowed to snowball from a simple seed, it identifies not just \textit{more} stakeholders, but \textit{better connected} ones—agents that bridge gaps between the core roles.

Consequently, practitioners do not need to invest heavily in curating exhaustive stakeholder lists. A simple seed list, combined with a dynamic network expansion process and betweenness-based prioritization, is sufficient to uncover high-quality, novel risks that outperform static nuanced lists.

\subsection{Prolific Expert and Rubric Validation}
\label{app:prolific}
\rev{To ensure high-quality annotations for our comparison, we recruited five external AI-domain experts via Prolific. We applied strict screening criteria: participants were required to have a professional background in AI ethics assessment, and pass both an initial comprehension check and embedded attention checks during the task. They were compensated above minimum wage.}

\rev{To validate our rubric dimensions, we computed the Pearson correlation among these scores across all evaluated risks (Figure \ref{app:prolific}). The low-to-moderate correlations among most of these metrics confirm that these dimensions capture distinct evaluative constructs rather than redundant signals (although as expected, Likelihood is more strongly correlated with Plausibility). Novelty is marginally negatively correlated with Likelihood and Uniqueness, while weakly positively correlated with all other measures.}

\clearpage

\begin{table*}[t]
\small
\centering
\caption{Effect of Seeding Strategy on Risk Prioritization (Small-World + Betweenness Centrality). Dynamic expansion (Snowballing) drastically improves performance over fixed stakeholders. Notably, snowballing from a small, generic set (Typical) slightly outperforms starting with a more specific set (Nuanced) on Combined Score (3.99 vs 3.94) and Novelty (3.48 vs 3.42), indicating that organic network growth effectively discovers critical bridging roles.}
\label{tab:seeding}
\begin{tabular}{llcc}
\toprule
\textbf{Initial Seed} & \textbf{Expansion Strategy} & \textbf{Combined} & \textbf{Novelty} \\
\midrule
Typical (5 generic stakeholders) & Fixed (No expansion) & 3.19 & 2.36 \\
Nuanced (30 specific stakeholders) & Fixed (No expansion) & 3.73 & 3.12 \\
\midrule
Typical (5 generic stakeholders) & Snowballing & \textbf{3.99} & \textbf{3.48} \\
Nuanced (30 specific stakeholders) & Snowballing & 3.94 & 3.42 \\
\bottomrule
\end{tabular}
%\vspace{-0.15in}
\end{table*}

\begin{table*}[t]
\centering
\small
\caption{\revision{Demographics of participants in the Chatbot Companion workshops for our first evaluation (blank slate human generated risks).}}
\begin{tabular}{l c}
\toprule
\textbf{\revision{Characteristic}} & \textbf{\revision{Number of Participants (\%)}} \\

\midrule
\textbf{\revision{Gender}} & \\
\revision{Men} & \revision{15 (33.3\%)} \\
\revision{Women} & \revision{30 (66.7\%)} \\

\midrule
\textbf{\revision{Age}} & \\
\revision{18--24} & \revision{8 (17.8\%)} \\
\revision{25--34} & \revision{23 (51.1\%)} \\
\revision{35--44} & \revision{8 (17.8\%)} \\
\revision{45+}  & 1 \revision{(2.2\%)} \\

\midrule
\textbf{\revision{Education}} & \\
\revision{No degree} & \revision{0 (0.0\%)} \\
\revision{Bachelor's degree} & \revision{8 (17.8\%)} \\
\revision{Master's degree} & \revision{23 (51.1\%)} \\
\revision{PhD} & \revision{9 (20.0\%)} \\

\midrule
\textbf{\revision{AI Expertise}} & \\
\revision{Basic} & \revision{5 (11.1\%)} \\
\revision{General} & \revision{7 (15.6\%)} \\
\revision{Knowledgeable} & \revision{17 (37.8\%)} \\
\revision{Expert} & \revision{12 (26.7\%)} \\

\midrule
\textbf{\revision{Field of Work/Study}} & \\
\revision{CS / AI / Data Science} & \revision{27 (60.0\%)} \\
\revision{Health \& Life Sciences} & \revision{6 (13.3\%)} \\
\revision{Physical Sciences} \& \revision{Engineering} & 2 (4.4\%) \\
\revision{Humanities \& Ethics} & \revision{3 (6.7\%)} \\
\revision{Business} & \revision{3 (6.7\%)} \\
\revision{Others} & \revision{4 (8.9\%)} \\

\midrule
\textbf{\revision{Total participants}} & \revision{45} \\

\bottomrule
\end{tabular}
\label{tab:chatbot_participants}
\end{table*}

\clearpage

\subsection{\revision{Risk Impact Assessment Reports}}
\label{appendix:reports}

\begin{figure*}[h!]
    \centering
    \includegraphics[width=0.8\linewidth]{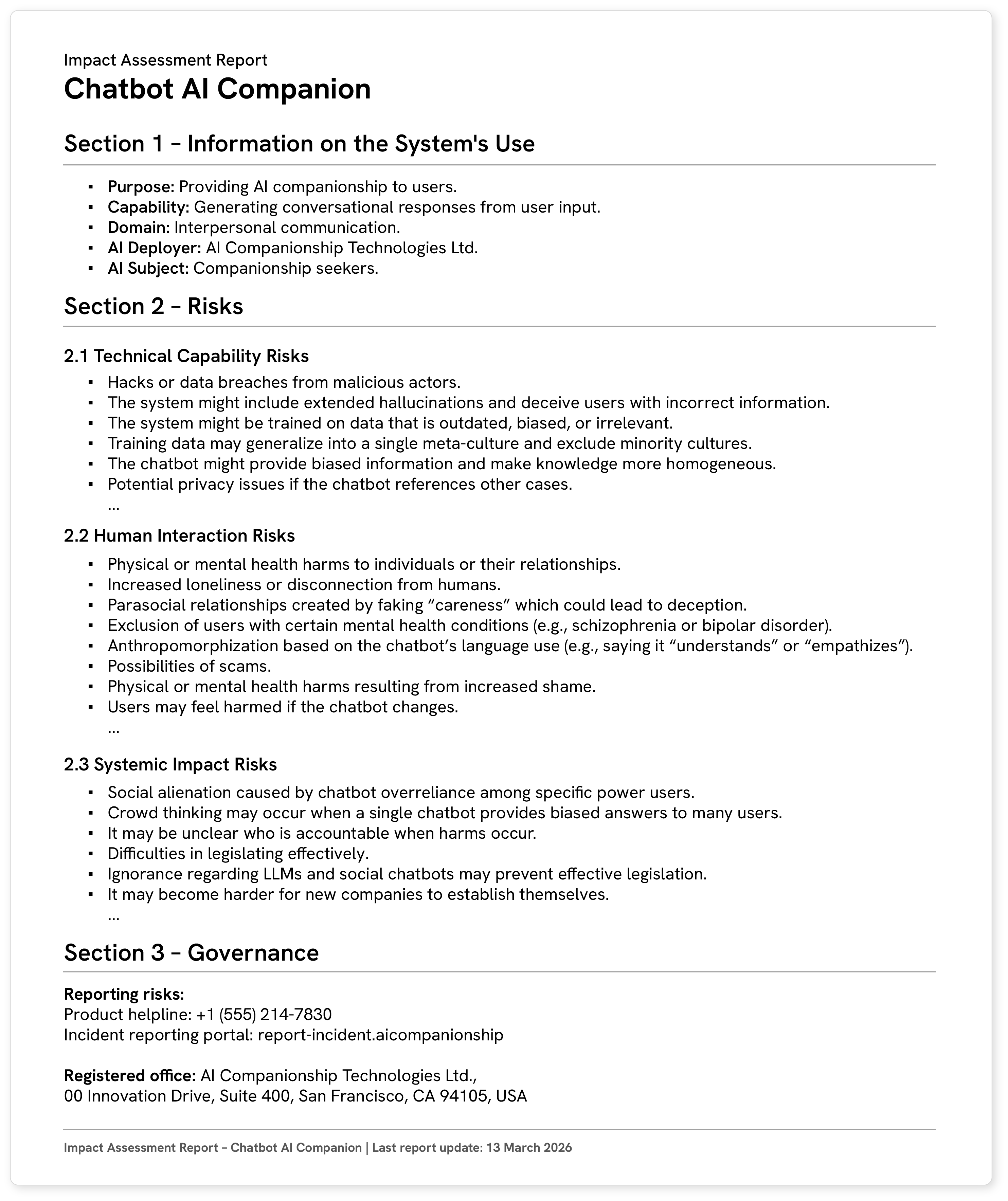}
    \caption{\revision{\textbf{Risk impact assessment report provided to participant groups in the control condition of the second evaluation.} The report contains risks generated by AI practitioners in blank-slate brainstorming workshops (Section \ref{sec:eval1_workshop}), organized into technical, human interaction, and systemic risk categories.}}
    \label{fig:report_baseline}
    %\vspace{-0.15in}
\end{figure*}

\begin{figure*}[t]
    \centering
    \includegraphics[width=0.8\linewidth]{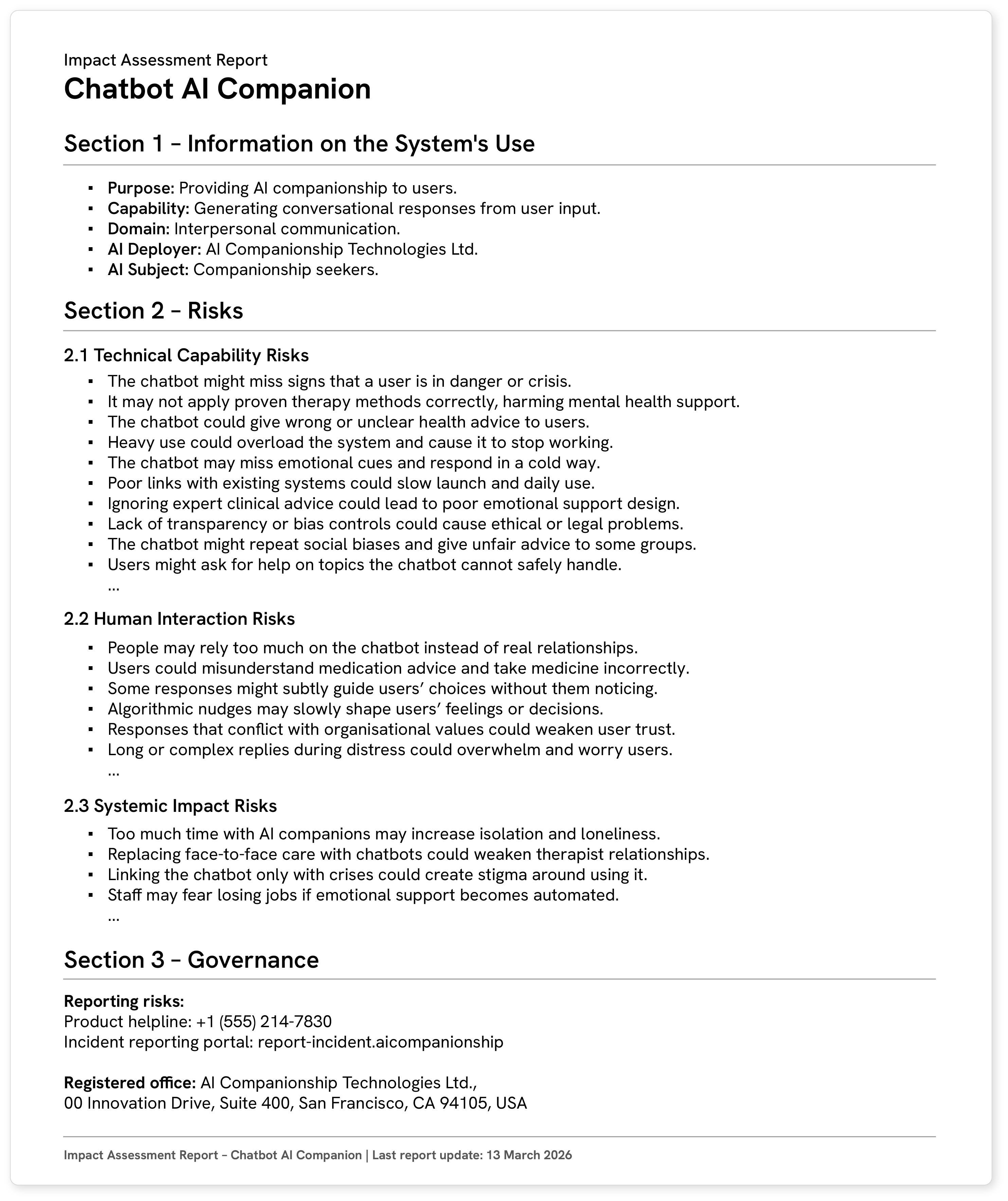}
    \caption{\revision{\textbf{Risk impact assessment report provided to participant groups in the treatment condition of the second evaluation.} The report contains risks generated by our networked stakeholder simulation framework (Section \ref{sec:networkagent}), organized into technical, human interaction, and systemic risk categories.}}
    \label{fig:report_treatment}
    %\vspace{-0.15in}
\end{figure*}

\begin{figure*}[t]
    \centering
    \includegraphics[width=0.6\linewidth]{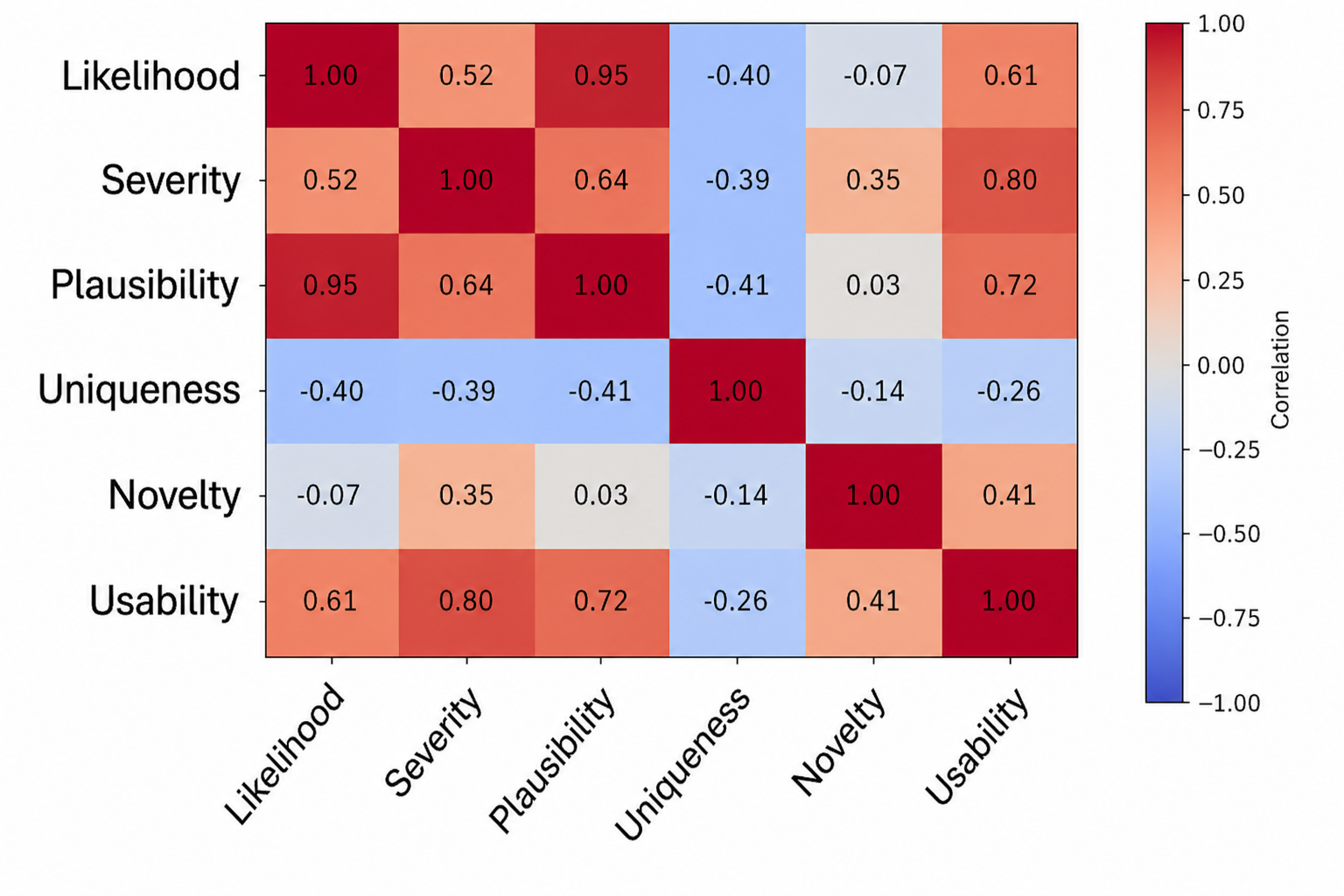}
    \caption{\rev{Correlation matrix of various rubric dimensions demonstrates that these dimensions capture distinct evaluative signals rather than redundant measures. Novelty is marginally negatively correlated with Likelihood and Uniqueness, while weakly positively correlated with all other measures.}}
    \label{fig:corelation}
\end{figure*}

\end{document}